\documentstyle[psfig,epsf,amssymb]{aa}
\begin{document}
\thesaurus{02.01.2, 02.08.1, 02.09.1, 11.01.2, 11.14.1}
\title{Accretion discs models with the $\beta$-viscosity prescription derived 
from laboratory experiments}
\author{Jean-Marc Hur\'e$^{1,2}$, Denis Richard$^{2,3}$ and
Jean-Paul Zahn$^3$}
\authorrunning{J.-M. Hur\'e, D. Richard \& J.-P. Zahn}
\titlerunning{Accretion discs models with the $\beta$-viscosity 
prescription}
\offprints{Jean-Marc.Hure@obspm.fr}
\institute{$^1$DAEC et UMR 8631 du CNRS, Observatoire de
Paris, Place Jules Janssen, 92195 Meudon
Cedex, France\\
           $^2$Universit\'e Paris 7 Denis Diderot, 2 Place Jussieu,
75251
Paris Cedex 05, France\\
           $^3$DASGAL et UMR 8633 du CNRS, Observatoire de
Paris, Place Jules Janssen, 92195 Meudon
Cedex, France}
\date{Received ; Accepted}

\maketitle

\begin{abstract}

We examine under which conditions one may apply, to steady state
keplerian accretion discs, the $\beta$-viscosity
prescription which has been derived from rotating shear flow experiments
($\nu = \beta \Omega R^2$, where $\Omega$
is the angular velocity at radius $R$ and $\beta$ is a constant
of order $10^{-5}$ (Richard \& Zahn 1999)).
Using a vertically averaged model, we show that this law may be suitable for
all three families of known systems: in young stellar objects,
evolved binary stars and Active Galactic Nuclei discs (except in their
outer gas pressure
dominated regions where turbulence becomes hypersonic). According
to the standard criterion for viscous stability, $\beta$-discs are always
stable throughout.  Using realistic opacities and
equation of state, we demonstrate that these discs
are thermally unstable in the temperature domain where
hydrogen recombines, when they are optically thick, and 
this could lead to limit cycle
behavior.
Radiation pressure dominated
regions are thermally stable, in contrast with $\alpha$-discs. This
results in a fully stable solution for the innermost parts of AGN discs.

\begin{keywords}
accretion, accretion discs | instabilities | hydrodynamics | galaxies:
active | galaxies: nuclei
\end{keywords}

\end{abstract}

\section{Introduction}

What causes the transport of matter and angular momentum in
accretion discs?
This question is the subject of a longstanding debate, but most of
those
working in this field tend now to agree that it has been answered,
since even a very weak magnetic field
leads to a
linear instability (Chandrasekhar 1961) which, in its fully
developed regime,
is able to provide the necessary stresses (Balbus \& Hawley 1991;
Hawley \&
Balbus 1991).  However, it is unlikely that this mechanism 
operates in cold
neutral discs
(e.g. Gammie \& Menou 1998). Besides, the issue which has not
been settled
yet is whether this magnetic instability is the only instability
responsible
for the ``anomalous viscosity'', or whether a purely hydrodynamic
instability,
generated by shear in differentially rotating discs, may not
be
of comparable or even higher efficiency.
Since a non-magnetic keplerian disc is
linearly stable, this hydrodynamic instability needs a finite
amplitude
perturbation to be triggered and laboratory experiments indicate
that it
appears only above Reynolds numbers of the order of $10^4$. Such
conditions are still
out of reach of the most ambitious numerical simulations;
therefore, these
cannot be used yet to prove or to disprove the occurrence of
shear
instability.

It is worth noting that the widely used $\alpha$-prescription
for
turbulent viscosity (Shakura \& Sunyaev 1973)
does not invoke any particular mechanism at the
origin of
turbulence: it is a simple parameterization which is tailored
to yield turbulent velocities that remain subsonic (or sub-alfv\'enic)
for $\alpha < 1$.  It is often asserted that the main success of the
$\alpha$-prescription
is the prediction of recurrent outbursts of dwarf novae
(e.g. Cannizzo 1993). However, if such eruptive phenomena are
indeed of thermal origin, as is commonly thought, it is likely that
many {\it
  ad-hoc} viscosities, not only the $\alpha$-prescription, could
work as well.

For these multiple reasons it makes
sense to investigate the basic properties of
accretion discs built with the alternate
$\beta$-prescription, which is observed
in laboratory  rotating shear flows, as explained below. This is our
intention in this paper. We first present in Sect. 2 some properties
of $\beta$-viscosity,
and examine in  Sect. 3 to which regimes
it may be applied, allowing for subsonic turbulent velocities in
geometrically thin, radiatively cooled accretion discs.
In Sect. 4 we study  the viscous and thermal stabilities of
$\beta$-discs,
using standard criteria, and draw our conclusions in the last section.
The Appendix contains the vertically averaged
equations for $\beta$-discs and the derivation
of the criterion of thermal instability.

\section{Comparing the $\alpha$ and $\beta$-viscosities}

The laboratory experiment which is the most pertinent to studies of
hydrodynamical instabilities in differentially rotating flows is the
Couette-Taylor
experiment: a fluid is sheared between two cylinders rotating at
different speeds. Only a few experiments have been run in the case
where the angular momentum increases outwards, as in a keplerian
disc, but from the torque measurements that are available, it
appears that the turbulent viscosity scales as
\begin{equation}
\nu_ {\rm t}  \propto R^3 \left|\frac{d\Omega}{dR}\right|,
\label{sup_nu}
\end{equation}
where $R$ is the distance from the rotation axis and $\Omega$ is the
angular velocity (Richard \& Zahn 1999). In a keplerian disc,
this relation is equivalent to
\begin{equation}
\nu_\beta = \beta \Omega R^2,
\label{eq:nubeta}
\end{equation}
which we shall call the $\beta$-prescription.

This prescription
 was originally proposed by Lynden-Bell \& Pringle (1974)
and some aspects of it have been discussed in later papers (De Freitas
Pacheco \&
Steiner 1976; Thompson et al. 1977; Lin \& Papaloizou 1980;
Williams 1980;
 Hubeny 1990). It has been recently revived by Biermann and Duschl
 (1998; see also Duschl, Biermann \& Strittmatter 2000) on the
grounds that
 the only relevant scales in a keplerian disc are the angular velocity
and the
 radius, since they contain all the information about the rotation and
the
 curvature of the flow. These authors make the sensible choice of
equating the parameter
 $\beta$ with the inverse of the critical Reynolds number ${\cal R}e$,
which they
 assume to be the same as in plane-parallel shear flows (i.e. ${\cal
R}e
 \approx 10^3$ and therefore $\beta \approx 10^{-3}$).

However, in the following we shall use a smaller value for
this
$\beta$-parameter, which is derived from the Couette-Taylor
experiment.
The only experimental data available, in the case where
angular momentum increases outwards, are those
obtained by Wendt (1933) and Taylor (1936),
in which the inner cylinder is at rest; from these one derives
$\beta \approx 10^{-5}$.
This value
is compatible with the critical gradient
Reynolds number which characterizes these experiments
(see Richard \& Zahn 1999):
\begin{equation}
{\cal R}e^*_{\rm c} = \frac{R^3}{\nu} \left|\frac{d\Omega}{dR}\right|
\approx 6
\times 10^5 ,
\end{equation}
where $\nu$ is the molecular viscosity of the fluid.
Clearly, new experiments are required to measure this parameter
in regimes which resemble more the keplerian flow; in the meanwhile,
until new results
are obtained, we shall adopt the value quoted above.

One important property of this $\beta$-viscosity is that it depends
only on the
radius in a keplerian disc, and does not involve local physical
conditions.
On the contrary, the standard
$\alpha$-prescription (Shakura \& Sunyaev 1973) depends on the local
values of the pressure scale
height $H$ and sound velocity $c_{\rm s}$ (taken in a broad sense --
it is the Alfv\'en velocity if magnetic pressure dominates):
\begin{equation}
\nu_\alpha = \alpha c_{\rm s} H .
\label{eq:nua}
\end{equation}
To emphasize the difference between the present prescription and
the standard law, we use the relation $c_{\rm s} = \Omega H$ which
expresses the hydrostatic equilibrium in a keplerian vertically
averaged disc, and rewrite Eq. (\ref{eq:nua}) as
\begin{equation}
\nu_\alpha = \alpha \Omega R^2  \left(\frac{H}{R}\right)^2_\alpha
= \nu_\beta \times  \frac{\alpha}{\beta} \times
\left(\frac{H}{R}\right)^2_\alpha .
\label{eq:nua2}
\end{equation}
We see that the ratio $\nu_\alpha/\nu_\beta$ depends on the
as\-pect ratio of the $\alpha$-disc. In
general, the quantity $(H/R)_\alpha$ is weakly sensitive to both
$R$ and
$\alpha$, but it depends strongly on the central mass and the accretion
rate, and is typically in
the range $0.001 - 0.1$  (e.g. Hur\'e 1998).

The $\alpha$-prescription has been
designed such that the Mach number is smaller than unity for 
 $\alpha \le 1$. In contrast, subsonic turbulence
 is not guaranteed with the $\beta$-prescription. This restriction on
subsonic turbulent velocities is dictated by the fact that supersonic
turbulence would be highly dissipative and therefore difficult to
sustain.
Moreover, the turbulent viscosity used here has been measured in a
liquid,
and its
application to compressible fluids can only be justified in the
subsonic range. Therefore one has to check whether the $\beta$-prescription
predicts subsonic turbulence in a given disc.
By making the reasonable assumption that the vorticity $v_{\rm
t}/\ell_{\rm
  t}$ of the turbulent eddies is of order $\Omega$ ($v_{\rm t}$ and
$\ell_{\rm
  t}$ denoting respectively the typical velocity and length scales of
  turbulence), and setting $\nu_\beta = v_{\rm t} \ell_{\rm t}$, the
Mach number in a $\beta$-disc is given by
\begin{equation}
{\cal M}_{\rm t} = \frac{v_{\rm t}}{c_{\rm s}} \simeq \sqrt{\beta}
\, \frac{\Omega R}{
c_{\rm s}} = \sqrt{\beta}
\left(\frac{H}{R}\right)_\beta^{-1}
\label{eq:macht}
\end{equation}
where $(H/R)_\beta$ is the aspect ratio of the $\beta$-disc. We
note that
${\cal M}_{\rm t} < 1$ provided  $H/R> \sqrt{\beta}$, which is not a
too
severe requirement with our low value of $\beta$ (the situation
would be more
problematic with $\beta=10^{-3}$, conflicting with the keplerian
assumption).

\section{Application range of the $\beta$-viscosity}

Let us now examine under which conditions the $\beta$-prescription
allows for subsonic turbulence in geometrically thin
accretion discs in active galactic nuclei (AGN),
evolved binary systems (EBS) and young stellar objects (YSO).
For this first exploration, we will be content with the usual
one-dimensional steady
state approach to compute the sound speed and the aspect ratio, the
vertical
structure being reduced to vertical averages.
The reader can find the relevant equation set in Appendix A.

In a stationary
keplerian disc characterized by a steady accretion rate $\dot{M}$,
the surface density $\Sigma$ is determined by the conservation of mass
and momentum:
\begin{equation}
\dot{M} = 3 \, \pi \nu \Sigma ,
\label{eq:angmom}
\end{equation}
ignoring a correcting factor which is negligible sufficiently far from
the center (see Frank et al. 1992).
Applying the $\beta$-prescription (\ref{eq:nubeta}), we find
\begin{eqnarray}
\Sigma & \simeq& 2.43 \times 10^5 \; \beta_{-5}^{-1}\, f_{edd} \,
x^{-1/2} \qquad {\rm
g.cm}^{-2}
\nonumber
\\
& \simeq& 92 \; \beta_{-5}^{-1} \, \dot{m}_{16} \, M_0^{-1/2} \,
r_{10}^{-1/2} \qquad {\rm g.cm}^{-2}
\nonumber
\\
 & \simeq& 70 \, \, \beta_{-5}^{-1} \, \dot{M}_{-8} \, M_0^{-1/2} \,
r_{\rm AU}^{-1/2} \qquad {\rm g.cm}^{-2}
\label{eq:sigma_all_value}
\end{eqnarray}
where $ f_{edd} \simeq 4.5 \; \dot{M}_0 \; M_{8}^{-1} $ is the Eddington
factor (around 0.1 typically),
$M = M_8 \times 10^8$ M$_\odot = M_0 $ M$_\odot$,
$\dot{M} = \dot{M}_0$ M$_\odot = \dot{m}_{16} \times 10^{16}$
g/s $= \dot{M}_{-8} \times
10^{-8}$ M$_\odot$/yr, $R= x \, R_{\rm S} = R_{10}
\times 10^{10}$ cm $=r_{\rm AU}$ AU ($R_{\rm S} = 2 GM/c^2$
being the Schwarzschild radius of the black hole), and $\beta_{-
5}=\beta
\times 10^{5}$. These dimensionless variables are well suited to
describe the three families of discs quoted above.
 Note that Eq. (\ref{eq:sigma_all_value}) is quite
general
and does not depend on the vertical structure.

Thermal equilibrium between radiative cooling and viscous heating
provides another equation.
Assuming that the disc is optically thick,
we have (Frank, King \& Raine 1992):
\begin{equation}
\frac{16 \, \sigma T^4}{3 \, \kappa_{\rm R} \Sigma} = \frac{3 \,
\Omega^2 \dot{M}}{8 \, \pi}
\label{eq:tc}
\end{equation}
where $\sigma$ is the Stefan constant and $\kappa_{\rm R}$ the Rosseland mean
opacity.

\subsection{Regimes compatible with subsonic turbulence}

In {\it gas pressure supported discs}, the hydrostatic equation
yields directly the mid-plane temperature $T$:
\begin{equation}
{P_{\rm g} \over \rho} = \Omega ^2 H^2 = {k T \over \mu m_{\rm H}} ,
\label{eq:hyd_gas}
\end{equation}
with the usual notations for the gas pressure $P_{\rm g}$,
the density $\rho$, the Boltzmann constant $k$, the molecular
weight $\mu$ and the mass of the hydrogen atom $m_{\rm H}$.
It then follows from
Eqs. (\ref{eq:macht}), (\ref{eq:sigma_all_value}),
(\ref{eq:tc}) and  (\ref{eq:hyd_gas})
that the turbulent Mach number in those
discs is
\begin{eqnarray}
{\cal M}_{\rm t} &\simeq& 3.1 \; \beta_{-5}^{5/8} \, \kappa_{\rm
R}^{-1/8} \mu^{1/2} \, f_{edd}^{-1/4} \, M_8^{1/8} x^{-1/16}
\nonumber
\\
 &\simeq&0.3 \; \beta_{-5}^{5/8} \, \kappa_{\rm R}^{-1/8}
\mu^{1/2} \, \dot{m}_{16}^{-1/4} \, M_0^{7/16} r_{10}^{-1/16}
\nonumber
\\
 &\simeq& 4 \times 10^{-2} \; \beta_{-5}^{5/8} \, \kappa_{\rm
R}^{-1/8} \mu^{1/2} \, \dot{M}_{-8}^{-1/4} \, M_0^{7/16} r_{\rm
AU}^{-1/16}  .
\label{eq:macht2}
\end{eqnarray}
 Note the relatively strong dependence of the Mach
number on the viscosity parameter and
its weak dependence on the opacity and specially the radius.

We conclude that the $\beta$-prescription is
certainly applicable to gas pressure dominated
accretion discs around forming stars, white dwarfs, neutrons
stars and stellar black holes, but not to such discs around
supermassive black holes, where this prescription would imply
supersonic turbulence.

The innermost part of accretion discs may be dominated  by
{\it radiation pressure} if
they reach a sufficiently high temperature; this occurs for instance in
standard
discs
surrounding black holes or compact objects. In such a case, the
hydrostatic equation is written
\begin{equation}
{P_{\rm r} \over \rho} = \Omega ^2 H^2 = {8 \sigma \over 3 c}
{T^4 H \over \Sigma} ,
\label{eq:hyd_rad}
\end{equation}
and therefore, according to (\ref{eq:tc}),
$H$ depends only on the accretion rate, and not on the viscosity
prescription:
\begin{equation}
H = {3 \kappa_{\rm R} \over 16 \pi c}  \dot{M} .
\end{equation}
Thus the Mach number is given by
\begin{eqnarray}
{\cal M}_{\rm t} &\simeq& 3 \times 10^{-3} \, \beta_{-5}^{1/2} \,
\kappa_{\rm
  R}^{-1}\,  f_{edd}^{-1} \, x
\nonumber
\\
&\simeq& 5 \times 10^{-2} \, \beta_{-5}^{1/2} \, \kappa_{\rm R}^{-1}\,
\dot{m}_{16}^{-1} \,  M_0 \, x
\end{eqnarray}
meaning that turbulence is expected to be subsonic in the
inner part of radiation pressure dominated discs.

\begin{figure} 
\psfig{figure=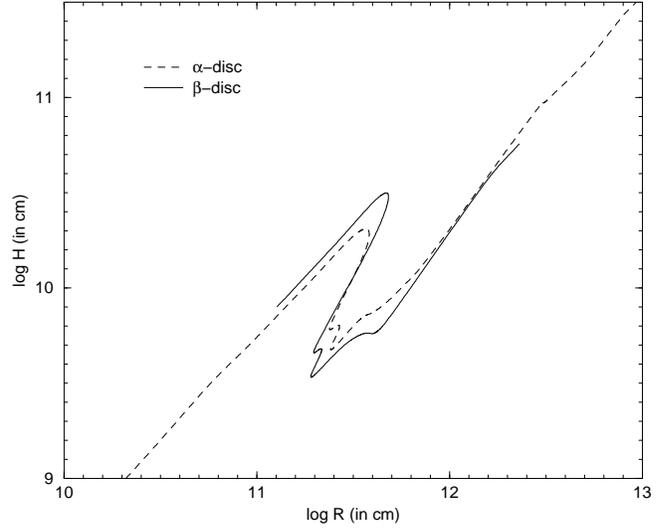,width=8.6cm}
\caption{Disc thickness versus radius for a $\beta$-disc
with $\beta=10^{-5}$
({\it solid line}) and for a standard disc with
 $\alpha=0.01$ ({\it dashed line}), for $=1$ M$_\odot$ and
$\dot{M}=10^{-8}$ M$_\odot$/yr.}
\label{fig:hr}
     \end{figure}

\subsection{Thin keplerian discs assumption}

In this paper, we restrict ourselves to keplerian discs, which are
necessarily
geometrically thin, i.e. $H/R \ll 1$. Eq. (\ref{eq:macht}) shows that the
turbulent Mach number increases as the aspect ratio decreases; therefore
the disc cannot be
thinner than $\sqrt{\beta} \times R$. One easily finds from
Eq. (\ref{eq:macht2}) that
\begin{eqnarray}
\left(\frac{H}{R}\right)_\beta & \simeq & 1.5 \times 10^{-3} \,
\kappa_{\rm R}^{1/8} \, \beta_{-5}^{-1/8}\, \mu^{-1/2} m_0^{-
7/16} \, \dot{m}_{16}^{1/4} \, r_{10}^{1/16}
\nonumber
\\
 & \simeq &  0.08 \, \kappa_{\rm R}^{1/8} \, \beta_{-5}^{-1/8}
\mu^{-1/2} \, m_0^{-7/16} \, \dot{m}_{-7}^{1/4} \, r_{\rm
AU}^{1/16}
\label{eq:hoverrbeta}
\end{eqnarray}
within the gas pressure dominated regions.
It is interesting again to make
the
comparison with standard discs. Using power law solutions of
$\alpha$-discs (Hur\'e 1998) and Eq. (\ref{eq:hoverrbeta}), we obtain
\begin{equation}
\frac{H_\beta}{H_\alpha} \simeq 0.9 \; \beta_{-5}^{1/8}
\alpha^{1/10} ,
\end{equation}
indicating that  $\beta$-discs globally should have about the same shape
as
standard $\alpha$-discs. This is confirmed
 in Fig.~\ref{fig:hr}, where we have plotted the disc
thickness
computed from vertically averaged equations with realistic
opacities and
equations of state, for the two
viscosity laws and for $M=1$ M$_\odot$ and $\dot{M}=10^{-8}$
M$_\odot$/yr
which could represent both a YSO disc and a EBS disc. 
In the innermost part of AGN, where the radiation pressure dominates,
the disc thickness is independent of the viscosity prescription, as
we have already pointed out, and therefore $H_\beta=H_\alpha$.
We thus conclude that, despite the flaring, $\beta$-discs are
expected to be thin and
keplerian, provided that the accretion rate does not reach too
large values. Note that $\alpha$-discs are subject to the same constraint.

\section{Stability of $\beta$-discs}

We shall now examine two kinds of instability that an accretion disc can
undergo: viscous instability which can lead to disc
fragmentation and thermal instability.

\subsection{Viscous stability}

\begin{figure} 
\psfig{figure=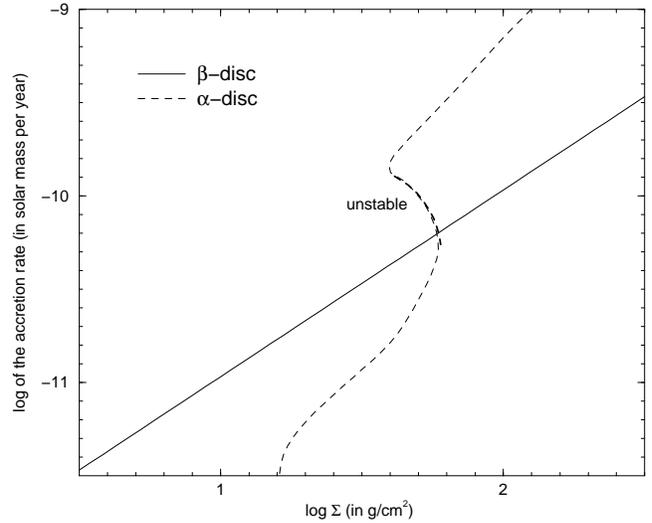,width=8.5cm}
\caption{Relation between the accretion
rate and the surface density, from vertical structure computations,
at $R=10^{10}$ cm from the center and with $M$=1 M$_\odot$,
using the $\alpha$-viscosity
prescription (with $\alpha=0.1$) and the $\beta$-prescription
($\beta=10^{-5}$).}
\label{fig:Sshapecuve}
\end{figure}

A general
condition for viscous stability is (Frank et al. 1992)
\begin{equation}
\left(\frac{\partial \ln \dot{M}}{\partial \ln
\Sigma}\right)_\Omega > 0 .
\label{eq:visccrit}
\end{equation}
In a standard $\alpha$-disc, the equilibrium relation $\dot{M}(\Sigma)$
takes the famous ``S''-shape
with the instability occuring on the intermediate branch of negative slope
(Cannizzo 1993).
On the contrary, according to Eqs. (\ref{eq:nubeta}) and (\ref{eq:angmom})
 $\beta$-discs are viscously stable everywhere, since
Ineq. (\ref{eq:visccrit}) is always
 fulfiled. We illustrate this important property in
Fig.~\ref{fig:Sshapecuve}, with two typical
$\dot{M}(\Sigma)$-relations obtained through vertical structure
computations for a $\beta$-disc and a $\alpha$-disc (Hur\'e, 2000).

\subsection{Thermal stability}

Thermal instability occurs when the energy balance between
radiative losses
and viscous heating can no longer be maintained after perturbation
of the
equilibrium state. The condition for stability is written (Pringle
1981):
\begin{equation}
\left(\frac{\partial \ln Q^+}{\partial \ln T}\right)_{\Sigma,
  \Omega} - \left(\frac{\partial \ln Q^-}{\partial \ln
T}\right)_{\Sigma,
  \Omega} < 0
\label{eq_etatherm}
\end{equation}
where $Q^+$ is the heating rate (right-hand-side of
Eq. (\ref{eq:tc})) and $Q^-$ is the cooling rate (with $Q^+=Q^-$ at
equilibrium). The above criterion is quite general
and can be applied to non standard viscosity prescriptions,
provided that the
vertical dynamical time scale $t_z \sim H/c_{\rm s}$ is smaller than the
thermal
time scale $t_{\rm th} \sim PH/Q^+$. The requirement of a
subsonic turbulence
discussed in the previous section automatically ensures that $t_z
\lesssim t_{\rm th}$ since we have the relation
\begin{equation}
\frac{t_{\rm th}}{t_z} \sim \frac{1}{ \beta } \left( \frac{H}{R}
\right)^2 = \frac{1}{{\cal M}^2_{\rm t}} .
\label{eq_tscales}
\end{equation}

It is easy to see that $Q^+ \propto \Omega^3 R^2 \Sigma$ with the
$\beta$-prescription and
so the thermal stability of a $\beta$-disc is only governed by
properties of the
cooling function. The situation is different in an $\alpha$-disc where
the heating rate $Q^+ \propto \Omega^3 H^2 \Sigma$
is implicitly a function of $T$ through $H$. To determine the slope of the
cooling function
$Q^-(T)$, we have used vertically averaged equations (see
Appendix A) and computed the term $(\partial \ln Q^- /
\partial \ln T)$
analytically (see Appendix B).

\begin{table*}
\caption{Quantity $(\partial \ln Q^- / \partial \ln T)_{\Sigma,
\Omega}$ and
  inferred thermal stability for ideal opacity sources (disregarding
self-gravitation). Values and comments in parenthesis concern the
$\alpha$-prescription. Various cases are
considered: $\tilde{\beta}=1$ ($\tilde{\beta}$ is the ratio of the gas
pressure to the total pressure); $\tilde{\beta}=\frac{1}{2}$ and
$\tilde{\beta}=0$ (radiative pressure only); Thompson scattering
only ($\kappa \equiv \sigma_{\rm T}$); free-free opacity only
($\kappa \equiv \kappa_{\rm ff} \propto \rho T^{-7/2}$); both
absorption processes (with $x = \frac{\kappa_{\rm
ff}}{\kappa_{\rm ff}+\sigma_{\rm T}} \le 1$).}
\begin{tabular}{l|l|cl|cl} \hline
$\beta$ & opacity & \multicolumn{2}{c|}{optically thick limit (see
Eq. (\ref{eq_etathick}))} & \multicolumn{2}{c}{optically thin limit
(see Eq. (\ref{eq_etathin}))} \\ \hline
&&&&& \\
$1$                      & $\sigma_{\rm T}$    & $-4$           & stable
\qquad ($-3$, stable) & $-4$ & stable \qquad ($-3$, stable)\\
                         & $\kappa_{\rm ff}$   & $-8$           & stable
\qquad ($-7$, stable) & $ 0$ & stable \qquad ($+1$, {\bf
unstable})\\
                         & $\sigma_{\rm T} + \kappa_{\rm ff}$ & $-4(1+x)
<
0$ & stable    & $-4(1-x) \le 0$ & stable\\

$\frac{1}{2}$            & $\sigma_{\rm T}$ & $-4$ & stable \qquad
($-\frac{2}{3}$, stable) & $-4$ & stable \qquad ($-\frac{2}{3}$,
stable)\\
                         & $\kappa_{\rm ff}$  & $-\frac{55}{6}$ & stable
\qquad ($-\frac{35}{6}$, stable) & $-\frac{1}{8}$ & stable \qquad
($+\frac{5}{8}$, {\bf unstable})\\
                         & $\sigma_{\rm T} + \kappa_{\rm ff}$ & $ -4 -
\frac{31}{6}x < 0$ \qquad & stable & $-4+\left( \frac{10-7x}{6+2x}
+\frac{7}{2} \right) x <0$ & stable\\

$0$                      & $\sigma_{\rm T}$ & $-4$ & stable \qquad
($+4$, {\bf unstable}) & $-4$ & stable \qquad ($+4$, {\bf
unstable})\\
                         & $\kappa_{\rm ff}$  & $-\frac{23}{2}$ & stable
\qquad ($-\frac{7}{2}$, stable) & $-\frac{1}{4}$ & stable \qquad
($+\frac{1}{4}$, {\bf unstable})\\
                         & $\sigma_{\rm T} + \kappa_{\rm ff}$ & $-4 -
\frac{15}{2}x < 0$ & stable & $-4+\left( \frac{8-7x}{2+2x}
+\frac{7}{2} \right) x <0$ & stable \qquad \\

\hline
\end{tabular}
\label{tab:epsilon}
\end{table*}

The resulting expression Eq. (\ref{eq_6bis}) may be used to
analyze the thermal stability  for very simple ideal cooling
functions:
electron scattering and free-free processes, as considered
already
by Piran (1978). We have
distinguished 4 cases: optically thick, optically thin,
gas pressure dominated and radiation pressure
dominated. This can easily be handled by suitable
settings in Eq. (\ref{eq_6bis}) (see Eqs. (\ref{eq_etathick}) and
Eq. (\ref{eq_etathin})). The results are summarized in
Tab.(\ref{tab:epsilon}). The remarkable point is that $\beta$-discs
are always
thermally stable for these ideal cooling mechanisms, which is in complete
agreement
with the conclusions drawn by Piran (1978). Note that the optically
thin gas pressure dominated regime
with free-free emission is marginally stable
whereas it is unstable with the standard viscosity law, as noticed by
Pringle,
Rees \& Pacholczyk (1973).

\begin{figure} 
\psfig{figure=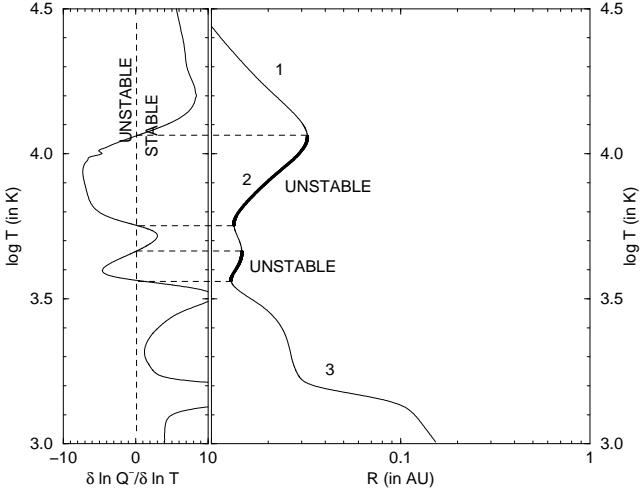,width=8.6cm}
\caption{{\it Right panel}: Temperature radial profile ({\it thin
line}) and
  thermally instable regions ({\it bold line}) in a
  "realistic" $\beta$-disc for $M=1$ M$_\odot$ and $\dot{M}=10^{-
8}$ M$_\odot$/yr. {\it Left panel}:
  the quantity $(\partial \ln Q^- / \partial \ln T)_{\Sigma,
\Omega}$ versus $T$
  computed from Eq. (\ref{eq_6bis}).}
    \label{fig:newinsyso}
     \end{figure}

\begin{figure} 
\psfig{figure=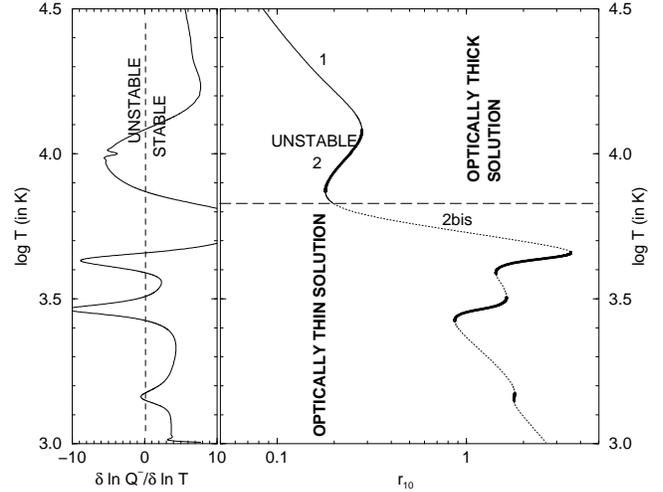,width=8.6cm}
\caption{Same legend as for Fig.~\ref{fig:newinsyso} but
  for $\dot{M}=10^{-12}$ M$_\odot$/yr typical of a quiescent EBS
disc. The optically
  thick/thin limit is reached at $r_{10} \simeq 0.2$ where $T
\simeq 6600$ K. The optical thickness $\tau$ goes through a
minimum for $T \simeq 4500$ K where $\tau \simeq 4 \times
10^{-4}$.}
    \label{fig:newinsebs}
     \end{figure}

The situation is quite different when  using
 realistic Rosseland and Planck opacities
as well as an accurate equation
of state. A typical result is displayed in Fig.~\ref{fig:newinsyso},
where we have considered
a disc surrounding a one solar mass central object and
$\dot{M}=10^{-8}$
M$_\odot$/yr. We find that the disc is thermally unstable
 when the midplane temperature is in the range $5000 - 9000$ K
(zone 2 on the figure), due to the steep rise in
the opacity. Like in a standard disc, this unstable regime
corresponds to the partial ionization of
hydrogen. A specially interesting case is displayed in
Fig. \ref{fig:newinsebs} where the accretion rate is
$\dot{M}=10^{-12}$ M$_\odot$/yr , typical for an EBS disc in
quiescence. There is a (multi-valued) optically thin
solution with $T \lesssim 6600$ K (zone 2bis in the figure). We
have checked
that this regime is also compatible with both $H/R \ll 1$ and ${\cal
M}_{\rm
  t}<1$. We see that the hottest branch of this
solution, which has already been investigated by Williams (1980), is
thermally stable.

Similar computations have been carried out for many input pairs
($M, \dot{M}$) corresponding to AGN, EBS and YSO discs and the
overall
conclusions are the following:
\begin{itemize}
\item the hot optically thick solution which extends down to the
central
  object (zone 1 in Figs. 3 and 4) is always thermally stable 
(this includes the radiative pressure dominated solution);
\item the region where hydrogen recombines (zone 2) is always
unstable if the disc is optically thick;
\item the hottest branch of the optically thin solution with $7000$
K $\gtrsim T \gtrsim 4000$ K
  and $dT/dR < 0$ (zone 2bis), when it exists, is always stable, even if
hydrogen recombination occurs;
\item the cold branches of the optically thin solution with $T
\lesssim 4000$
  K and $dT/dR >0$ are always unstable;
\item the outermost parts (zone 3) are globally stable. Thin
unstable zones can be present.
\end{itemize}

\section{Conclusion}

In this paper, we have examined some properties of thin accretion discs
built with the $\beta$-prescription for the turbulent viscosity,
which has been derived from the transport of angular momentum
observed in differentially  rotating laboratory
flows (Richard \& Zahn 1999).
This prescription may be applied whenever it predicts a turbulent velocity
that is subsonic, and we have verified that, with the value 
we have taken for $\beta$ (i.e. $10^{-5}$), this condition is
fulfiled in most discs,
around stellar objects or massive black holes,
except in the outer part of AGN discs where gas pressure dominates.
Only radiatively cooled discs have been surveyed; it would be interesting
to investigate ADAF-type solutions with this prescription.
Let us recall that our conclusions have been obtained here with
vertically averaged models,
and it is plausible that a more realistic 2D-model would extend
somewhat the domain of applicability
of the $\beta$-prescription (Hur\'e \& Richard 2000).

Although the $\beta$-viscosity and the
 $\alpha$-viscosity may have comparable magnitudes,
they have very different effects both on the structure and on the
stability of steady  keplerian discs.
An important property of $\beta$-discs is that they
are viscously stable and do not tend to fragment
into concentric rings, contrary to $\alpha$-discs.
Also, unlike $\alpha$-discs, they are thermally stable for ideal
cooling
processes, such as Thomson scattering and free-free absorption, as
was shown by Piran (1978). However, with more realistic opacities
and equation of state, models of $\beta$-discs contain a
thermally unstable region corresponding to the recombination of hydrogen,
much like standard $\alpha$-discs.
It remains to be seen with time-dependent models
whether this instability can lead, as we suspect,
to a limit cycle behavior.

\begin{acknowledgements}
The authors wish to thank the referee for his competent remarks and 
suggestions.
\end{acknowledgements}

\bigskip

\appendix

\section{Vertically averaged equations for a steady state keplerian
$\beta$-disc}

In a steady state keplerian accretion disc where the viscosity is
given by
Eq.(\ref{eq:nubeta}), the mass accretion rate $\dot{M}$ is linked to
the
surface density $\Sigma$ by the relation
\begin{equation}
\dot{M} = 3 \, \pi \beta \Omega R^2 \Sigma,
\label{eq_angmom}
\end{equation}
where $\Omega$ is the keplerian angular velocity and $R$ the
radius. The surface density is  $\Sigma = 2 \rho H$, with $\rho$ being
the mass density and $H$ the total pressure scale height, which is
determined from the equation of hydrostatic equilibrium
\begin{equation}
\frac{P}{\rho H} = \Omega^2 H (1 + \zeta) \quad \hbox{with} \quad 
\zeta = {4 \pi G \rho \over \Omega^2} ,
\label{eq_hydrostatic}
\end{equation}
where $\zeta$ represents the effect of self-gravity.
The total pressure $P$ 
is the sum of the gas pressure $P_{\mathrm g}$ and of the
radiation pressure $P_{\mathrm r}$:
\begin{equation}
 P = P_{\mathrm g} + P_{\mathrm r} = \frac{P_{\mathrm
g}}{\tilde{\beta}} .
\label{eq_pression}
\end{equation}
We draw  $P_{\mathrm g}$ from the perfect gas equation
\begin{equation}
P_{\mathrm g} = \frac{\rho k T}{\mu m_{\mathrm H}},
\label{eq_pgas}
\end{equation}
where $T$ is the midplane temperature and $\mu$ the mean
mass per particle
in units of the proton mass. The disc temperature in the equatorial
plane is
fixed by the balance $Q^+=Q^-$ between viscous heating
(Frank, King \& Raine, 1992)
\begin{equation}
Q^+ = \frac{3\, \Omega^2 \dot{M}}{8\, \pi}
\label{eq_heating}
\end{equation}
and radiative losses (Hubeny 1990)
\begin{equation}
Q^- = \frac{8\, \sigma T^4}{3\, \kappa_{\mathrm R}  \rho H +
\frac{2}{\kappa_{\mathrm P} \rho H} +2 \sqrt{3}},
\label{eq_cooling}
\end{equation}
where $\kappa_{\rm R}$ and $\kappa_{\rm P}$ are respectively the Rosseland
and Planck mean
opacities (in cm$^2$.g$^{-1}$). The general
expression for the radiative pressure is then    \begin{equation}
P_{\mathrm r}=\frac{Q^-}{c}\left( \frac{1}{2} \kappa_{\rm R} \rho
H+\frac{1}{\sqrt{3}}\right) .
\label{eq_prad}
\end{equation}
\onecolumn

Note that in the optically thick limit  the radiation
pressure tends
to its LTE value (labeled with an asterisk)
 \begin{equation}
P_{\mathrm r}=\frac{4 \sigma T^4}{3 c} \equiv P_{\rm r}^* .
\label{eq_pradLTE}
\end{equation}

The above equation set can in fact be reduced to a system of two
non linear algebraic equations with $\Sigma$ as the unknown. From
Eqs.(\ref{eq_angmom})-(\ref{eq_prad}), the
first equation is
\begin{equation}
P^* - \pi G  \Sigma^2 - \frac{\pi^6 \beta^6 G^4 M^4}{256 \,
\dot{M}^6} \frac{\Sigma^8}{\rho} - \frac{16 \, \sigma T^4}{9 \, c
\kappa_{\rm R} \kappa_{\rm P} {\mathcal S}(\Sigma) } = 0,
\label{eq_general1}
\end{equation}
where the function ${\cal S}(\Sigma)$ is
\begin{equation}
{\mathcal S}(\Sigma) = \frac{\Sigma^2}{2} + \frac{2
\Sigma}{\sqrt{3} \kappa_{\rm R}} +\frac{4}{3 \kappa_{\rm R}
\kappa_{\rm P}} .
\label{eq_bigs}
\end{equation}
The second equation is obtained from Eqs.(\ref{eq_angmom}),
(\ref{eq_heating})
and (\ref{eq_cooling})
\begin{equation}
\frac{\sigma \dot{M}^5 T^4}{3^{14} \, \pi^{11} \, \kappa_{\rm
R}} = G^4 M^4  \beta^6 {\mathcal S}(\Sigma) \Sigma^5 .
\label{eq_useful1}
\end{equation}

We can follow the method
described in Hur\'e (1998) to solve the system of Eqs.(\ref{eq_general1})
and (\ref{eq_useful1}). Note that the solution is valid in all
cases: optically thick or thin, pressure dominated by gas or radiation,
self-gravitating or not.

\subsection{The optically thick limit}

If we set ${\mathcal S}(\Sigma) = \frac{1}{2} \Sigma^2$ in
Eqs.(\ref{eq_general1}) and (\ref{eq_useful1}), and omit the last term
of Eq.(\ref{eq_general1}) we recover the optically thick limit ($\tau \gg 1$),
and there is a single equation to solve:
\begin{equation}
\dot{M}^{12/7}+\lambda_1 \dot{M}^{2/7}+\lambda_0 = 0 .
\label{eq_mdotthick}
\end{equation}
The coefficients $\lambda_0$ and $\lambda_1$ are respectively
\begin{equation}
\lambda_0 = 16 \left( \frac{\pi^{5/6} \beta \sigma M^{2/3}}{3
\sqrt{G} \kappa_{\rm R}}\right)^{6/7} \frac{T^{24/7}}{\rho}
\label{eq_lamda0} \quad \hbox{and} \quad
\lambda_1 = - \frac{3}{4} \left( \frac{3 \kappa_{\rm R} \pi
\beta^6 \sqrt{\pi G} M^4}{\sigma}\right)^{2/7} \frac{P^*}{T^{8/7}}
\label{eq_lamda1} .
\end{equation}
Note that the solution of Eq.(\ref{eq_mdotthick}) must satisfy $\tau
\gg 1$.

\subsection{The optically thin limit}

In the optically thin limit ($\tau \ll 1$), ${\mathcal S}(\Sigma) =
4 /  3  \kappa_{\rm R} \kappa_{\rm P}$ and the last term of
Eq.(\ref{eq_general1}) is
identical to $P_{\rm r}^*$. There is also a single equation to solve,
namely

\begin{equation}
\left[4 \, \pi G \rho + \left( \frac{16 \sigma \kappa_{\rm P}}{3
\beta G^{2/3} M^{2/3}}T^4 \right)^{6/5} \right] \dot{M}^2 
- \frac{3 \, k \rho^2}{\mu m_{\rm H} T^{3/5}} \left(\frac{6 \pi^5
\beta^6 G^4 M^4}{\sigma \kappa_{\rm P}}\right)^{2/5} = 0
\label{eq_mdotthin}
\end{equation} 
and its solution must be such that $\tau \ll 1$.


\section{Criterion for the thermal instability}

Setting $\tau= \kappa_{\rm R} \Sigma + 8 / 3 \, \kappa_{\rm
P} \Sigma + 4/\sqrt{3}$, and differentiating
Eq.(\ref{eq_cooling}) while keeping $\Sigma$ constant, we obtain
\begin{eqnarray}
\left(\frac{\partial \ln Q^-}{\partial \ln T}\right)_{\Sigma,\Omega}
=  4 & -& \frac{1}{\tau} \left[ \tau_{\rm R} b_{\rm R} -
\frac{8  b_{\rm P} }{3 \, \tau_{\rm P}} \right]
 - \frac{1}{\tau} \left[ \tau_{\rm R} a_{\rm R} -
\frac{8  a_{\rm P}}{3 \, \tau_{\rm P}} \right]  \left(\frac{\partial \ln
\rho
}{\partial \ln T}\right)_{\Sigma,\Omega},
\label{eq_2}
\end{eqnarray}
where $\tau_{\rm R} = \kappa_{\rm R} \Sigma$ and $\tau_{\rm P}
= \kappa_{\rm P} \Sigma$ are the total Rosseland and Planck
optical thicknesses respectively, and
$$a_{\rm R}=\left(\frac{\partial \ln \kappa_R}{\partial \ln
\rho}\right)_{T},
\qquad
 b_{\rm R}=\left(\frac{\partial \ln \kappa_R}{\partial \ln
T}\right)_{\rho},
\qquad
 a_{\rm P}=\left(\frac{\partial \ln \kappa_P}{\partial \ln
\rho}\right)_{T} , \qquad
 b_{\rm P}=\left(\frac{\partial \ln \kappa_P}{\partial \ln
T}\right)_{\rho} .
$$
Note that this expression does not depend on the actual viscosity
prescription. For Thompson scattering, we have simply $a_{\rm R} =
b_{\rm R} =
a_{\rm P} = b_{\rm P} =0$, and for free-free opacity, $a_{\rm R} =
a_{\rm
  P}=1$ and $b_{\rm R} = b_{\rm P} = -\frac{7}{2}$. For a
combination of both
sources, we have $a_{\rm R} = a_{\rm P}=x$ and $b_{\rm R} =
b_{\rm P} =
-\frac{7}{2}x$ where $x \equiv \kappa_{\rm
ff} / (\kappa_{\rm ff}+\sigma_{\rm T})$.

To compute the derivative $\left(\partial \ln \rho / \partial \ln T
\right)_{\Sigma,\Omega}$, we use the two different 
expressions  of the total pressure at the midplane. First, by
differentiation of Eq.(\ref{eq_hydrostatic}), keeping $\Sigma$ and
$\Omega$ constant, we have
\begin{equation}
\left( \frac{ \partial \ln P }{\partial \ln T} \right)_{\Sigma,\Omega}
=  - \frac{1}{1+\zeta} \left( \frac{ \partial \ln \rho }{\partial \ln
T} \right)_{\Sigma,\Omega} .
\label{eq_3}
\end{equation}
Note that $(\partial \ln \rho / \partial \ln T)_{\Omega,\Sigma}$ is
always
negative. Second, from Eq.(\ref{eq_pression}) which is written
\begin{eqnarray}
P = P_{\rm g} + \frac{Q^-}{c} \left( \frac{\tau_{\rm R} }{4} +
\frac{1}{\sqrt{3}}\right)
\end{eqnarray}

\noindent and setting
 $\chi_\rho^{\rm g}=\left(\partial \ln P_{\rm g} / \partial
\ln T \right)_{\rho}$, $\chi_T^{\rm g}=\left( \partial \ln P_{\rm g} /
\partial \ln \rho \right)_T$ and $\tilde{\beta}=P_{\rm g}/P$,
we have
\begin{equation}
\left( \frac{ \partial \ln P }{\partial \ln T} \right)_{\Sigma,\Omega}
= \left[ \tilde{\beta} \chi_T^{\rm g} +   \left(1-
\tilde{\beta}\right)  a_{\rm R} \frac{\tau_{\rm R}}{\tau_{\rm R} +
\frac{4}{\sqrt{3}}} \right] \left( \frac{ \partial \ln \rho }{\partial
\ln T} \right)_{\Sigma,\Omega}
 + \tilde{\beta}\chi_\rho^{\rm g} +
\left(1-\tilde{\beta}\right) \frac{\tau_{\rm R}}{\tau_{\rm R} +
\frac{4}{\sqrt{3}}} b_{\rm R} + \left(1-\tilde{\beta} \right)
\left(\frac{\partial \ln Q^- }{\partial \ln T}\right)_{\Sigma,\Omega} .
\label{eq_5}
\end{equation}
 Combining now Eqs.(\ref{eq_3}) and (\ref{eq_5}), we find
\begin{equation}
-\left( \frac{ \partial \ln \rho }{\partial \ln T}
\right)_{\Sigma,\Omega} \left[ \frac{1}{1+\zeta} + \tilde{\beta}
\chi_T^{\rm g} + \left(1-\tilde{\beta}\right)  a_{\rm R}
\frac{\tau_{\rm R}}{\tau_{\rm R} + \frac{4}{\sqrt{3}}} \right] =
\tilde{\beta}\chi_\rho^{\rm g} + \left(1-\tilde{\beta}\right)
\frac{\tau_{\rm R}}{\tau_{\rm R} + \frac{4}{\sqrt{3}}} b_{\rm R} +
\left(1-\tilde{\beta} \right) \left(\frac{\partial \ln Q^- }{\partial
\ln
T}\right)_{\Sigma,\Omega} .
\label{eq_6}
\end{equation}
Finally, from Eqs.(\ref{eq_2}) and (\ref{eq_6}) and after
rearrangement
\begin{eqnarray}
\left(\frac{\partial \ln \rho }{\partial \ln
T}\right)_{\Sigma,\Omega} & = & \left[4(1-\tilde{\beta}) -(1-
\tilde{\beta}) \frac{\tau_{\rm R} b_{\rm R} - \frac{8 \, b_{\rm P}
}{3 \, \tau_{\rm P}}}{\tau} + \tilde{\beta} \chi_\rho^{\rm g} + (1-
\tilde{\beta}) \frac{\tau_{\rm R}}{\tau_{\rm R} + \frac{4}{\sqrt{3}}}
b_{\rm R} \right]
\nonumber
\\
& \times & \left[ (1-\tilde{\beta}) \frac{\tau_{\rm R} a_{\rm R} -
\frac{8 \, a_{\rm P} }{3 \, \tau_{\rm P}}}{\tau} - \left(
\frac{1}{1+\zeta} + \tilde{\beta} \chi_T^{\rm g} +  \left(1-
\tilde{\beta}\right) a_{\rm R} \frac{\tau_{\rm R}}{\tau_{\rm R} +
\frac{4}{\sqrt{3}} } \right)\right]^{-1}
\label{eq_6ter}
\end{eqnarray}
and so Eq.(\ref{eq_2}) becomes
\begin{eqnarray}
\left(\frac{\partial \ln Q^- }{\partial \ln T}\right)_{\Sigma,\Omega}
& = & \left[ \tilde{\beta} \chi_\rho^{\rm g} + \left(1-
\tilde{\beta}\right) \frac{\tau_{\rm R}}{\tau_{\rm R} +
\frac{4}{\sqrt{3}}} b_{\rm R} +  \left( \frac{1}{1+\zeta} +
\tilde{\beta} \chi_T^{\rm g} +  \left(1-\tilde{\beta}\right) a_{\rm R}
\frac{\tau_{\rm R}}{\tau_{\rm R} + \frac{4}{\sqrt{3}} } \right) \left(
\frac{4 \tau - \tau_{\rm R} b_{\rm R} + \frac{8 \, b_{\rm P} }{3 \,
\tau_{\rm P}} }{\tau_{\rm R} a_{\rm R} - \frac{8 \, a_{\rm P} }{3 \,
\tau_{\rm P}}}\right)  \right]
\nonumber
\\
& \times & \left[\frac{\tau }{ \tau_{\rm R} a_{\rm R} - \frac{8 \,
a_{\rm P} }{3 \, \tau_{\rm P}}} \left( \frac{1}{1+\zeta} +
\tilde{\beta} \chi_T^{\rm g} +  \left(1-\tilde{\beta}\right) a_{\rm R}
\frac{\tau_{\rm R}}{\tau_{\rm R} + \frac{4}{\sqrt{3}} } \right) +
\tilde{\beta} -1\right]^{-1}
\label{eq_6bis}
\end{eqnarray}

It is then easy to derive from Eq.(\ref{eq_6bis})  the following
asymptotic expressions:
\begin{itemize}

\item in the optically thick limit ($\tau \sim \tau_{\rm R}$ and
$\tau_{\rm P} \gtrsim \tau_{\rm R} \gg 1$)
\begin{equation}
-\left(\frac{\partial \ln Q^- }{\partial \ln T}\right)_{\Sigma,\Omega}
= b_{\rm R} - 4 - a_{\rm R} \frac{\tilde{\beta} \chi^{\rm g}_{\rho} +
4(1-\tilde{\beta})}{ \tilde{\beta} \chi^{\rm g}_T + \frac{1}{1+\zeta}} ,
\label{eq_etathick} 
\end{equation}

\item in the radiative pressure dominated limit ($\tilde{\beta} \sim
0 $)
\begin{equation}
-\left(\frac{\partial \ln Q^- }{\partial \ln T}\right)_{\Sigma,\Omega}
= b_{\rm R} - 4 - 4 a_{\rm R} (1+\zeta) ,
\end{equation}

\item in the gas pressure dominated limit ($\tilde{\beta} \sim 1 $)
\begin{equation}
-\left(\frac{\partial \ln Q^- }{\partial \ln T}\right)_{\Sigma,\Omega}
=
 b_{\rm R} - 4 - a_{\rm R} \frac{1 + \zeta}{2+\zeta} ,
\end{equation}

\item in the optically thin limit ($\tau \sim \frac{8}{3 \, \tau_{\rm
P}}$ and $\tau_{\rm P} \ll 1$)
\begin{equation}
-\left(\frac{\partial \ln Q^- }{\partial \ln T}\right)_{\Sigma,\Omega}
= - b_{\rm P} - 4 + a_{\rm P} \frac{\tilde{\beta} \chi^{\rm g}_{\rho}
+4(1-\tilde{\beta})+ (1-\tilde{\beta})b_{\rm P}}{
(1-\tilde{\beta})a_{\rm P}
+\tilde{\beta} \chi^{\rm g}_T + \frac{1}{1+\zeta}}  .
\label{eq_etathin}
\end{equation}

\end{itemize}

\end{document}